\documentclass[11pt,a4paper]{article}
\usepackage[british]{babel}
\usepackage{enumerate,eepic}
\usepackage[T1]{fontenc}
\usepackage[dvips]{graphicx}
\usepackage{psfrag,times}
\usepackage{geometry}
\usepackage{amssymb}
\newcommand{\anymode}[1]{\ifmmode{#1}\else\mbox{$#1$}\fi}
\newcommand{\biblioitem}[1]{\par \frenchspacing
           \vbox{\noindent \hangindent=2em \hangafter=1{#1}}
           \vskip 2ex}

\def\biblioitem{\par \vskip 1ex \noindent \hangindent=2em \hangafter=1}
\def\linea#1{\ifhmode\hfill\break\fi\hbox to \hsize{#1}}

\newcommand{\inv}{^{-1}}

\def\vc{v\kern -0.1em .c.\relax}

\newcommand{\dfrac}[2]{\displaystyle{\frac{#1}{#2}}}

\newcommand{\ignore}[1]{}

\newcommand{\indep}{\perp\kern-0.5em\perp}

\newcommand{\Real}{\mathbb{R}}

\newcommand{\T}{^{\top}}

\newcounter{Definition}

\newcounter{Proposition}

\newtheorem{definition}[Definition]{Definition}

\newtheorem{proposition}[Proposition]{Proposition}
\def\oggi{\number\day
  \space\ifcase\month\or
    gennaio\or febbraio\or marzo\or aprile\or maggio\or giugno\or
    luglio\or agosto\or settembre\or ottobre\or novembre\or dicembre\fi
  \space\number\year}

\newcommand{\SN}{\mathop{\rm SN}\nolimits}

\renewcommand{\d}{\,\textrm{d}}

\title{
       On the canonical form of scale mixtures of skew-normal
       distributions
      }
\author{
        \textsc{Antonella Capitanio} \\
       {Department of Statistical Sciences,  University of Bologna}  \\
       \texttt{antonella.capitanio@unibo.it}
  }
\date{First version 3rd June 2008\\
      Second revision 31st March 2010\\
      Last revision 1st July 2012}
\begin{document}
\maketitle
\begin{abstract}

The canonical form of scale mixtures of multivariate skew-normal
  distribution is defined, emphasizing its role in summarizing some key properties
  of this class of distributions.  It is also shown that the canonical form corresponds to an affine invariant co-ordinate system as defined in Tyler \emph{et} al. (2009), and a method for obtaining the linear transform that converts a scale mixture of multivariate skew-normal
  distribution into a canonical form  is presented. Related results, where the particular case
  of the multivariate skew $t$ distribution is considered in greater
  detail, are the general expression of the Mardia indices of multivariate skewness and
  kurtosis and the reduction of dimensionality in calculating the
  mode. 

\end{abstract}
\vspace{3ex} \noindent\emph{Key-words:~} affine invariance, kurtosis, Mardia indices
of multivariate skewness and kurtosis, scale mixtures of normal
distributions, skewness, skew-normal distribution, skew $t$
distribution. \clearpage
\section{Introduction}

The Gaussian model plays a central role in statistical modelling;
nevertheless the need of flexible multivariate parametric models
which are able to represent departure from normality is testified
by the increasing weight of the literature devoted to this issues
during the last decade. Departure from normality can take place in
different ways, such as multimodality, lack in central symmetry,
excess or negative excess of kurtosis.  The present paper focuses on the last two
features, considering the class of distribution generated by scale mixtures of the $d$-dimensional skew-normal random variables defined by Azzalini and DallaValle (1996). 

The class of scale mixtures of skew-normal distributions includes parameters to regulated either skewness or kurtosis, and reduces to the class of scale mixture of normal distributions when the skewness parameter vanishes. Finally, the skew-normal distribution is recovered when the mixing distribution corresponds to a random variable that is equal to one with probability 1. 
Among the members of this family, whose general form has been firstly introduced by Branco and Dey (2001), the skew $t$ distribution is the one that has received the greatest attention;  it corresponds to the case where the mixing distribution is $W^{-1/2}$, where $W$ is a $Gamma(\nu/2,\nu/2)$ random variable.  Azzalini and Capitanio (2003) developed a systematic study of its main probabilistic properties as well as statistical issues, however some aspects have been left unexplored, like the expression of suitable indices of multivariate skewness and kurtosis and a formal proof of unimodality. The usefulness of the skew $t$ distribution has been explored in different applied problems. Azzalini and Genton (2008) proposed and discussed the use of the multivariate skew $t$ distribution as an attractive alternative to the classic robustness approach, and Walls (2005), Meucci (2006) and Adcock (2009), among others, adopted this model to represent relevant features of financial data. Another member which has been studied in some details is the multivariate skew-slash distribution, defined by  Wang and Genton (2006), which is obtained when the mixing distribution is $U^{-1/q}$, where $U$ is a uniform distribution on the interval $(0, 1)$ and  $q$ is a real parameter greater than zero.
 
This paper introduces the definition of a canonical form associated to scale
mixtures of skew normal distribution, which generalizes the analogous one
introduced in Azzalini and Capitanio (1999) for the multivariate
skew-normal distribution. The motivation is its suitability in allowing a simplified representation of some relevant features which are
shared by all the members of the class of scale mixtures of skew-normal distributions. In fact the components of the canonical form are such that all but one is symmetric: the skewed component summarizes the skewness of the distribution as a whole, leading to consistent simplifications in obtaining summary measures of the data shape.
For instance, compact general expressions for the indices
of multivariate skewness and kurtosis defined by Mardia (1970,
1974) for the entire class of scale mixtures of skew-normal distributions are obtained. 
It will be also shown that a data transformation leading to a canonical form generates an affine invariant co-ordinate system of the kind   defined and discussed in Tyler \emph{et} al. (2009) in connection with a general method for exploring multivariate data.  

\section{The skew-normal distribution and its canonical form}  \label{s:sect-2}
The multivariate skew-normal distribution has been defined in
Azzalini and Dalla Valle (1996). The parameterization  adopted in the
present paper is the one introduced by Azzalini and
Capitanio~(1999), that have further explored the properties of this
family. 

A $d$-dimensional variate $Z$ is said to have a skew-normal distribution
if its density function is
\begin{equation}\label{f:msn-dens}
f(z) = 2 \phi_d(z - \xi;\Omega)\Phi(\alpha\T \omega\inv (z-\xi))  \qquad (z \in \Real^d),
\end{equation}
where $\phi_d(z;\Omega)$ denotes the $d$-dimensional normal density with zero mean and full rank covariance 
matrix $\Omega$, $\Phi$
is the $N(0,1)$ distribution function, $\xi \in \Real^d$ is the
location parameter, $\omega$ is a diagonal matrix of scale
parameters such that $\bar\Omega =\omega\inv \Omega \omega\inv$ is
a correlation matrix, and $\alpha \in \Real^d$ is a shape parameter which
regulates departure from symmetry. Note that when $\alpha =0$ the
normal density is recovered. A random variable with
density~(\ref{f:msn-dens}) will be denoted by
$\SN_d(\xi,\Omega,\alpha)$. The skew-normal distribution shares many
properties with the normal family, such as closure under
marginalization and affine transforms, and $\chi^2$ distribution
of certain quadratic forms. See Azzalini and Capitanio (1999) for
details on these issues. For later use we recall that the mean
vector and the covariance matrix of $Z$ are
\begin{equation} \label{f:sn-mean-var}
    \mu=\xi+\left(\frac{2}{\pi}\right)^{1/2}\omega\delta  \qquad \mathrm{and} \qquad
    \Sigma=\Omega - \frac{2}{\pi}\omega\delta\delta\T \omega,
\end{equation}
where
\begin{equation}  \label{f:delta}
    \delta=\frac{1}{(1+\alpha\T\bar\Omega\alpha)^{1/2}}\bar\Omega\alpha
\end{equation}
is a vector whose elements lie in the interval
$(-1,1)$. From (\ref{f:delta}) we have also
\begin{equation}  \label{f:alpha}
\alpha=\frac{1}{(1- \delta\T \bar\Omega\inv
\delta)^{1/2}}\bar\Omega\inv \delta.
\end{equation}
It is important to note that the shape parameter of a marginal
component of $Z$ is in general not equal to the corresponding
component of $\alpha$. More specifically, when $Z$ is partitioned
as $Z=(Z_1,Z_2)\T$ of dimension $h$ and $d-h$, respectively, the
expression of the shape parameter of the marginal component $Z_1$
is given by
\[
  \bar\alpha_1 = \frac{\alpha_1 + (\bar\Omega_{11})\inv \bar\Omega_{12}\alpha_2}{(1+\alpha_2\T \bar\Omega_{22\cdot
  1}\alpha_2)^{1/2}},
\]
where $\bar\Omega_{22\cdot 1}=\bar\Omega_{22} - \bar\Omega_{21}
(\bar\Omega_{11})\inv \bar\Omega_{12}$, and $\bar\Omega_{ij}$ and
$\alpha_i$, for $i,j=1,2$, denotes the elements of the corresponding
partitions of $\bar\Omega$ and $\alpha$, respectively. On the
contrary, the entries of the vector $\delta$ after marginalization
are obtained by extracting the corresponding components of the original
parameter.

Azzalini and Capitanio (1999, Proposition~4) introduced a
canonical form associated to a skew-normal variate, via the
following result.
\begin{proposition} \label{p:sn-cf}
Let $Z \sim SN_d(\xi, \Omega, \alpha)$ and consider the affine non
singular transform $Z^*=(C\inv P)\T \omega\inv(Z - \xi)$ where
$C\T C=\bar\Omega$ and $P$ is an orthogonal matrix having the
first column proportional to $C\alpha$.  Then $Z^*\sim
SN(0,I_d,\alpha_{Z^*})$, where
$\alpha_{Z^*}=(\alpha_*,0,\ldots,0)\T$ and $\alpha_* = (\alpha\T
\bar\Omega \alpha)^{1/2}$. 
\end{proposition}

The above authors called the variate $Z^*$ a canonical form of $Z$.
With respect to the original definition, and without loss of
generality, here it is assumed that the non-zero element of the
shape vector $\alpha$ is the first one. The above result can be
easily verified by applying Proposition~3 of Azzalini and
Capitanio (1999). Furthermore, using their Propositions 5 and 6 it
is immediate to see that $Z_1^*\sim SN_1(0,1,\alpha_*)$ while the
remaining components of $Z^*$ are $N_1(0,1)$ variates, and that in addition
the components of $Z^*$ are mutually independent. Finally, it is remarked
that the linear transform leading to a canonical form is not
unique.

Azzalini and Capitanio (1999) underlined how this transformation plays a role analogous
to the one which converts a multivariate normal variable into a spherical
form. Motivated by the expressions they obtained for the
indices of multivariate skewness and kurtosis defined by Mardia
(1970), they also highlighted the role of $\alpha_*$ as a quantity
summarizing the shape of the distribution. In fact the two indices
are
\begin{eqnarray}\label{f:sn-g1}
\gamma_{1,d}&=&
\beta_{1,d}=\left(\frac{4-\pi}{2}\right)^2\left(\frac{2\alpha_*^2}{\pi + (\pi-2)\alpha_*^2} \right)^3\\
\label{f:sn-g2} \gamma_{2,d} &=&
\beta_{2,d}-d(d+2)=2(\pi-3)\left(\frac{2\alpha_*^2}{\pi +
(\pi-2)\alpha_*^2} \right)^2,
\end{eqnarray}
and they depends on $\alpha$ and $\Omega$ only via $\alpha_*$.

As an additional comment, note that by comparing expressions (\ref{f:sn-g1}) and (\ref{f:sn-g2})
with the corresponding ones for a univariate skew-normal distribution
(see Azzalini 1985, sect. 2.3), and taking into account
(\ref{f:alpha}), it turns out that the values of the multivariate skewness and
kurtosis indices are equal to those of the corresponding univariate indices for
a skew-normal distribution having shape parameter equal to
$\alpha_*$. In this sense, the canonical form is characterized by
one component absorbing the departure from normality of the whole
distribution.

Notice also that on using expression (\ref{f:delta}),  the marginal shape parameter $\delta$
associated to the skewed component of a canonical form turns out to be $\delta_*=(\delta\T
\Omega\inv \delta)^{1/2}$. Because of the one-to-one
correspondence between these two quantities, it makes no difference
which one is used as summary quantity.

Some results contained in Tyler \emph{et} al. (2009) allows to provide new insight into the role of a canonical transformation $Z^*$. These authors introduced a general method for exploring multivariate data, based on a particular invariant co-ordinate system, which relies on the eigenvalue-eigenvector decomposition of one scatter matrix relative to another. The canonical transformation $Z^*$ turns out to be an invariant co-ordinate system transformation with respect to the scatter matrices $\Omega$ and $\Sigma$, and taking into account the results of Section~3 of Tyler \emph{et} al. (2009), a method to obtain a matrix $H$ such that $Z^*=H\T (Z-\xi)$ can be explicitly stated. 
\begin{proposition} \label{p:sn-ics}
Let $Z \sim SN_d(\xi, \Omega, \alpha)$, and define $M = \Omega^{-1/2}\Sigma\Omega^{-1/2}$, where $\Omega^{1/2}$ is the unique positive definite symmetric square root of $\Omega$,  and $\Sigma$ is the covariance matrix of $Z$. Let $Q \Lambda Q\T$ denote the  spectral decomposition of $M$. Then the transform 
\[
Z^* = H\T(Z-\xi),
\]
where $H=\Omega^{-1/2}Q$, converts $Z$ into a canonical form. Moreover, $Z^*=H\T(Z-\xi)$ is an invariant co-ordinate system transformation based on the simultaneous diagonalization of the scatter matrices $\Omega$ and $\Sigma$.
\end{proposition}
\emph{Proof.} Consider the simultaneous diagonalization of the scatter matrices $\Omega=E[(Z-\xi)^2]$ and $\Sigma$, and let  $\Omega^{-1/2}$ denote the unique positive definite symmetric square root of $\Omega$. Following Tyler \emph{et} al. (2009, Section~3), a matrix $H$ such that $H\T\Omega H = I_d$ and $H\T\Sigma H= \mathrm{diag}(\lambda_1, \ldots,\lambda_d)$ turns out to be $\Omega^{-1/2} Q$, where $\lambda_1 \leq \lambda_2 \leq \ldots \leq \lambda_d$ are the eigenvalues of $\Omega\inv\Sigma$, or equivalently of $M=\Omega^{-1/2}\Sigma\Omega^{-1/2}$, and where the $i$th column of the $d\times d$ orthogonal matrix $Q$ is the normalized eigenvector of $M$ corresponding to the $i$th smallest eigenvalue. Furthermore, the $i$th column of $H$ is the eigenvector of $\Omega\inv\Sigma$ corresponding to the $i$th smallest eigenvalue of $\Omega\inv\Sigma$. The transform $H\T Z$ corresponds to an invariant co-ordinate system, as defined in Tyler \emph{et} al. (2009, p. 558). After some straightforward algebra, the eigenvalues of $\Omega\inv\Sigma$ turn out to be $1$, with multiplicity $d-1$, and $1-(2/\pi)\delta_*^2$, and the associated eigenspaces are the orthogonal complement of the subspace spanned by $\omega\delta$, and the subspace spanned by $\omega\inv\alpha$, respectively. This fact implies that the first row of $H\inv$ is proportional to $\omega\delta$, while the last $d-1$ rows lie in the orthogonal complement of the subspace spanned by $\omega\inv \alpha$. On using the expressions for the parameters of a linear transformation of a skew-normal variate given in Azzalini and Capitanio (1999, p. 585), the distribution of $Z^*=H\T (Z-\xi)$ is $SN(0,I_d, H\inv\omega\inv\alpha)$; taking into account the structure of the matrix $H\inv$ and the equality $H\inv\Omega\inv (H\inv)\T = I_d$ we obtain $H\inv\omega\inv\alpha = (\alpha_*, 0,\ldots, 0)\T$, and hence the variate $Z^*$ corresponds to the canonical form of $Z$. \emph{QED}

The proof of Proposition~\ref{p:sn-ics} contains a description of the structure of the matrix $H$, which it is shown to have one column proportional to $\omega\inv\alpha$ and the remaining ones belonging to the orthogonal complement of the subspace spanned by $\omega\delta$. This result implies that the projection $ \alpha\T\omega\inv Z$ captures all the skewness and the kurtosis of the joint distribution, whereas by projecting $Z$ onto the orthogonal complement of the subspace spanned by $\omega\delta$ independent $N(0,1)$ variates are obtained.

Since a matrix $H$ converting a skew-normal variate to its canonical form can be obtained through the simultaneous diagonalization of a pair of scatter matrices different from $\Omega$ and $\Sigma$, it is expected that when two scatter matrices, $V_1$ and $V_2$ say, are such that they become diagonal when the variate $Z$ is in canonical form, then Proposition~\ref{p:sn-ics} will continue to be valid if the matrices $\Omega$ and $\Sigma$ are replaced by $V_1$ and $V_2$. An example of such matrices will be given in Proposition~\ref{p:sn-ics-sigmakappa} at the end of Section~\ref{s:mardia}.

\section{Scale mixtures of skew-normal variates and their canonical form}
In this section a canonical form analogous to the one introduced
for the skew-normal distribution is defined for scale mixtures of
skew-normal distributions, and some properties are given.
\subsection{Scale mixtures of skew-normal variates}
Scale mixtures of skew-normal distributions have been considered in Branco and Dey (2001). This class of distributions contains the corresponding class of scale mixture of normal distribution and the skew-normal distribution as proper members, allowing to model a wide range of shapes.
A scale mixture of skew-normal distributions is defined as follows.
\begin{definition} \label{d:sm-sn}
Let $Y = \xi + \omega SZ$, where $Z \sim \SN_d(0, \bar\Omega,
\alpha)$ and $S>0$ is an independent scalar random variable.  Then
the variate $Y$ is a scale mixture of skew-normal distributions,
with location and scale parameters $\xi$ and $\omega$,
respectively.
\end{definition}
Note that, when $\alpha = 0$, $Y$ reduces to the corresponding
scale mixture of $N_d(0,\bar\Omega)$ distributions. 

The $m$th order moments of $Y$ can be calculated by differentiating the
moment generating function given in Branco and Dey (2001,
expression 4.1). An alternative and simpler way to obtain moments
is to follow the scheme used by Azzalini and Capitanio (2003,
expression (28)) for the moments of the skew $t$ distribution,
which arises when $S=W^{-1/2}$, and $W$ is a
$Gamma(\frac{1}{2}\nu,\frac{1}{2}\nu)$ random variable.
Specifically, assuming that $\xi=0$ and $\omega=I_d$, by
exploiting the stochastic representation given in
Proposition~\ref{d:sm-sn} we obtain
\begin{equation} \label{f:sm-sn-moment}
   E(Y^{(m)})=E(S^{m})E(Z^{(m)}),
\end{equation}
where $Y^{(m)}$ denotes a moment of order $m$. Note that to use
this formula only the knowledge of $m$th order moments of $S$ and $Z$ is
required.

An appealing property of an $SN_d(0, \Omega, \alpha)$ variate is
that the distribution of its any even functions is equal to the
one obtained by applying the same even function to a
$N_d(0,\Omega)$ variate. This fact can be easily seen by
considering Proposition~2 in Azzalini and Capitanio (2003) and
noting that the skew-normal distribution belongs to the broader
class of distribution generated by perturbation of symmetry which
the proposition is concerned with. As a corollary it follows from (\ref{f:sm-sn-moment}) that
even order moments of $Y-\xi$ are equal to those of the corresponding
scale mixture of normal distributions.
On using (\ref{f:sm-sn-moment}) and taking into
account (\ref{f:sn-mean-var}), the mean vector and the covariance
matrix of $Y$ are
\begin{equation} \label{f:sm-sn-meanvar}
E(Y) = \xi + E(S)\sqrt{\frac{2}{\pi}}\delta \qquad \mathrm{and} \qquad var(Y) =
E(S^{2})\bar\Omega - E(S)^2 \frac{2}{\pi}\delta \delta\T,
\end{equation}
in agreement with those obtained by Branco and Dey (2001).

Scale mixtures of skew-normal are models capable to take into
account for both skewness and kurtosis, and it is important to
have available the expressions of measures of these two features.
The next proposition introduces the expression of the Pearson
indices of skewness and kurtosis for the univariate case; the
multivariate case will be considered later, as the introduction
of the canonical form of $Y$ allows to cope with the problem in a
simpler manner.
\begin{proposition} \label{p:sm-sn-g1g2}
Let $Y = \xi + \omega SZ$, where $Z \sim \SN_1(0, 1, \alpha)$ and
$S>0$ is a scalar random variable. Then, provided that the moments
up to order three or up to order four of $S$ exist, the expressions of the skewness and
excess of kurtosis indices $\gamma_1$ and $\gamma_2$ are  
\begin{eqnarray*}   \label{f:gamma-1} 
  \gamma_1 &=& \beta_1 \,\, = \, \,\sigma_Y^{-3}\left(\frac{2}{\pi}\right)^{1/2}\left[E(S)^3\frac{4}{\pi}-E(S^{3})\right]\delta^3+\\
  \nonumber
           & &  -\sigma_Y^{-3}3\left(\frac{2}{\pi}\right)^{1/2}\left[E(S)E(S^{2})-
                   E(S^{3})\right]\delta,  \qquad  \mathrm{and}\\
  \gamma_2 &=& \beta_2-3 \,\, = \, \,\sigma_Y^{-4}\Bigg\{\frac{8}{\pi}\left[E(S)E(S^{3})-\frac{3}{\pi}E(S)^4\right]\delta^4+\\
  \nonumber
           & & -\frac{24}{\pi}\left[E(S)E(S^{3})-E(S^{2})E(S)^2\right]\delta^2+3\left[E(S^{4})-E(S^{2})^2
           \right]\Bigg\},
\end{eqnarray*}
where $\sigma_Y^{2}=var(Y)$.
\end{proposition}
\noindent \emph{Proof.} Since the two indices are location and
scale invariant, the case were $\xi=0$ and $\omega=1$ will be
considered. The third and the fourth cumulants of $Y$ required 
to compute $\gamma_1$ and $\gamma_2$ are functions of the
first four non central moments of $Y$, which in turn, taking into
account (\ref{f:sm-sn-moment}), depends on the
corresponding moments of $Z$. The first moment of $Z$ is given in
(\ref{f:sn-mean-var}), and taking into account that $Z^2 \sim
\chi_1^2$ (see Azzalini 1985, property H) the second and the
fourth ones are equal to 1 and 3, respectively. Finally, by
deriving the moment generating function of the scalar skew-normal
distribution given in Azzalini (1985, p. 174), the third moment of
$Z$ turns out to be $3(2/\pi)^{1/2}\delta - (2/\pi)^{1/2}\delta^3$.
After some algebra the result follows. \emph{QED}

Note that when $\delta =0$ the variate $Y$ is a scale mixture of
$N(0,1)$, so that the index $\gamma_1$ becomes zero and
$\gamma_2=\sigma_Y^{-4}3\left[E(S^{4})-E(S^{2})^2\right]$ measures
the excess of kurtosis of $Y$. When $S$ is degenerate and $S=1$,
the expressions of the two indices for the skew-normal
distribution are recovered. When $S$ is the inverse of the square
root of a $Gamma(\frac{1}{2}\nu,\frac{1}{2}\nu)$ random variable,
$Y$ follows a scalar skew $t$ distribution, and the two indices
coincide with those given in Azzalini and Capitanio (2003, p.
382).

\subsection{The canonical form of scale mixtures of skew-normal distributions}
The canonical form for scale mixtures of skew-normal distributions
is defined in the following way.
\begin{definition}  \label{d:cf_sm-sn}
Let $Y = \xi +  \omega SZ$, where $Z \sim \SN_d(0, \bar\Omega,
\alpha)$ and $S>0$ is an independent scalar random variable. The
variate $Y^*=(C\inv P)\T \omega\inv(Y - \xi)=SZ^*$, where the
matrices $P$ and $C$ are as in Proposition~\ref{p:sn-cf}, will be
called a canonical form of $Y$.
\end{definition}
From the above definition, it is straightforward to see that Proposition~\ref{p:sn-ics} can be extended to scale mixtures of skew-normal variates, that is, the linear transform $Y^*=H\T (Y-\xi)$, where $H$ is defined as in Proposition~\ref{p:sn-ics}, converts $Y$ into a canonical form.

The next proposition states some properties of $Y^*$.
\begin{proposition} \label{p:cf-properties}
Under the settings of Definition~\ref{d:cf_sm-sn}, the following
facts hold.
\begin{itemize}
    \item[(i)] Only the first univariate component of $Y^*$
    can be skewed. More specifically, $Y_1^*$ is a scale mixture of an
    $SN_1(0,1,\alpha_*)$ variate, where $\alpha_* = (\alpha\T
    \bar\Omega\alpha)^{1/2}$, and its mean and variance are
    \[
    \mu_*=E(S)(2/\pi)^{1/2}\delta_*, \qquad \sigma_*^2=E(S^{2})-(2/\pi)E(S)^2\delta_*^2, 
    \]
    respectively, where $\delta_*=(\delta\T\bar\Omega\inv\delta)^{1/2}$.
    The remaining components are identically distributed scale
    mixtures of $N_1(0,1)$ distributions, that is, symmetric about zero random variables
    with variance $\sigma^2=E(S^{2})$.
    \item[(ii)] The $d$ components of $Y^*$ are uncorrelated.
    \item[(iii)] The non zero elements of the set of moments $E(Y^{*(3)})$
    are
    \[
    E(Y_1^{*3})=E(S^{3})(2/\pi)^{1/2}\delta_*(3-\delta_*^2)
    \] 
    and
    \[
     E(Y_1^*Y_i^{*2})=E(S^{3})\sqrt{2/\pi}\delta_*, \qquad i=2,\ldots,d.
     \]
    \item[(iv)] The non zero elements of the set of moments $E(Y^{*(4)})$ are 
    \[
    E(Y_i^{*4})=3E(S^{4}), \qquad (i=1, \ldots, d),
    \]
    \[
      E(Y_i^{*2}Y_j^{*2})=E(S^{4}), \qquad (j=1, \ldots, d, \, i\neq j).
    \]
\end{itemize}
\end{proposition}
\emph{Proof.} $(i)$ By definition $Y_i^* = SZ_i^*$; the result
follows taking into account that $Z_1^* \sim SN_1(0,1,\alpha_*)$
whilst the last $d-1$ components of $Z^*$ are $N(0,1)$. The
expressions for the means and the variances can be obtained by
(\ref{f:sm-sn-meanvar}) taking into account (\ref{f:delta}).
$(ii)$ Using (\ref{f:delta}) the vector $\delta$ associated to
$Y^*$ becomes $(\delta_*,0,\ldots,0)\T$, where $\delta_*=(\delta\T
\bar\Omega\inv\delta)^{1/2}$; taking into account the expression
of $var(Y)$ given in (\ref{f:sm-sn-meanvar}), we see that
$Cov(Y^*_i,Y^*_j)=0$. $(iii)$--$(iv)$ From
expression~(\ref{f:sm-sn-moment}) we have
$E(Y^{*(m)})=E(S^{m})E(Z^{*(m)})$. The result follows taking into
account that the components of $Z^*$ are mutually independent and
the expressions of their moments. \emph{QED}

The above results show that the main features of the canonical form of the skew-normal
distribution are preserved when a scale mixture is considered. In fact only the first component is skewed, and the
influence of the parameters $\Omega$ and $\alpha$ is completely
summarized by quantity $\alpha_*$, or equivalently by $\delta_*$.
Independence among the components is replaced by a zero correlation, as expected since scale mixture of normal distribution themselves does not allow to model independence between components. 

\section{Mardia indices of multivariate skewness and kurtosis}  \label{s:mardia}
The canonical form of $Y$ can lead to dramatic simplification in
calculating quantities which are invariant or equivariant with
respect to invertible affine transformations. This is the case, for instance, of the Mardia indices of multivariate skewness and kurtosis and of the mode. In this section the Mardia indices will be considered, while the latter issue will be developed in the next section. 

Given a $d$-dimensional random variable $Y$, the Mardia indices of
multivariate skewness and excess of kurtosis are defined as
follows
\begin{eqnarray*} \label{d:mardia-inds}
\gamma_{1,d}&=& \; \beta_{1,d} \; = \;
\sum\limits_{ijk}\sum\limits_{i'j'k'}\sigma^{ii'}\sigma^{jj'}\sigma^{kk'}
    \mu_{i,j,k}\mu_{i'j'k'},\\
\gamma_{2,d}&=& \beta_{2,d} - d(d+2)\; = \;
E\left\{\left[(Y-\mu)\T\Sigma\inv(Y-\mu)\right]^2\right\} - d(d+2),
\end{eqnarray*}
where $\mu$ and $\Sigma$ denote the mean vector and the covariance
matrix of $Y$, respectively,
$\mu_{i,j,k}=E\left[(Y_i-\mu_i)(Y_j-\mu_j)(Y_k-\mu_k)\right]$,
and $\sigma^{ii'}$ denotes the $(i,i')$th entry of $\Sigma\inv$.\\
\begin{proposition} \label{p:mardia-d}
Consider the scale mixture of skew-normal distribution $Y = \xi +
\omega S Z$, where $Z \sim \SN_d(0, \bar \Omega, \alpha)$. Then the Mardia
indices of multivariate skewness and excess of kurtosis of $Y$ are, provided that the involved
moments of $S$ exist
\begin{eqnarray}   \label{f:gamma-2-d}
\nonumber
 \gamma_{1,d} &=& (\gamma_1^*)^2 + \frac{3(d-1)}{\sigma_*^2E(S^{2})^2}\left[E(S^{3})-E(S)E(S^{2})\right]^2\frac{2}{\pi}\delta_*^2,\\
\nonumber
 \gamma_{2,d} &=& \beta_2^* + (d-1)(d+1)E(S^{2})^{-2}E(S^{4})+ \\
 \nonumber
  & & \hskip-8pt
   +\frac{2(d-1)}{E(S^{2})\sigma_*^2}\Bigg\{E(S^{4}) +[E(S)^2E(S^{2})-2E(S)E(S^{3})]\frac{2}{\pi}\delta_*^2\Bigg\}-d(d+2),
 \end{eqnarray}
where, using a self explanatory notation, the quantities
$\gamma_1^*$, $\beta_2^*$, $\delta_*$ and $\sigma_*^2$ refer to
the component $Y_1^*$ of the canonical form associated to $Y$.
\end{proposition}
\emph{Proof.} In the proof some symbols introduced in
Proposition~\ref{p:cf-properties} will be used. Since
$\gamma_{1,d}$ and $\gamma_{2,d}$ are invariant with respect to
invertible affine transforms, the canonical form $Y^*$ will be
considered in place of $Y$. From $(i)$ of
Proposition~\ref{p:cf-properties} we know that the last $d-1$
components of $Y^*$ are symmetric about zero; a first implication
is that $\mu_{1,1,k}=0$ for any $2\leq k \leq d$. In addition,
taking into account $(iii)$, it follows that $\mu_{i,j,k}=0$ for
any choice of $i$, $j$ and $k$ in $\{2,\ldots, d\}$. From $(ii)$
we have $\sigma^{jj'}=0$ for any $j \ne j'$, and consequently
$\gamma_{1,d}$ reduces to
\[
\dfrac{(\mu_{1,1,1})^2}{\sigma_*^6}+\dfrac{3}{\sigma_*^{2}\sigma^{4}}\sum_{i=2}^d
\mu_{1,i,i}^2.
\]
Finally, by expressing $\mu_{1,i,i}$ in terms of
non central moments and by applying (\ref{f:sm-sn-moment}), the
first equality is proved.

Let us denote by $\mu_{i,j,k,l}$ the generic entry of the fourth
order central moment of $Y^*$; taking into account $(i)$ and
$(ii)$ of Proposition~\ref{p:cf-properties} we have
\begin{eqnarray*}
\beta_{2,d} &=&
E\left\{\left[\frac{(Y_1^*-\mu_*)^2}{\sigma_*^2}+\sum_{i=2}^d
\frac{Y_i^{*2}}{\sigma^2}\right]^2\right\}\\
 &=& \frac{\mu_{1,1,1,1}}{\sigma_*^4} +
\sum_{i=2}^d\frac{\mu_{i,i,i,i}}{\sigma^4} +
2\sum_{i=2}^{d-1}\sum_{j=3}^d\frac{\mu_{i,i,j,j}}{\sigma^4} +
2\sum_{i=2}^d\frac{\mu_{1,1,i,i}}{\sigma_*^2 \sigma^2},
\end{eqnarray*}
where the expressions of $\mu_{i,i,i,i}$ and $\mu_{i,i,j,j}$ for
$i$ and $j$ greater than 1 are given in $(iv)$ of
Proposition~\ref{p:cf-properties}, and that of $\mu_{1,1,i,i}$ can
be obtained with the aid of (\ref{f:sm-sn-moment}). After some
algebra the second equality follows. QED

This result shows that, if $Y$ is a scale mixture of skew-normal
distributions, then $\gamma_{1,d}$ and $\gamma_{2,d}$ depend on
the shape of $S$, and on the underlying skew-normal variate only via
the scalar quantity $\alpha_*$, or equivalently $\delta_*$,
reinforcing its role of a summary quantity of the distribution
shape.

By comparing these expressions with the corresponding ones of the
skew-normal distribution, given by (\ref{f:sn-g1}) and
(\ref{f:sn-g2}), respectively, we can observe that they have a
different structure. In particular, when a scale mixture of
skew-normal distributions is considered, the two indices do not
coincide with their univariate version evaluated with respect to
the marginal distribution of the only skewed component of the
variate in canonical form.

It could be of interest to highlight the structure of $\beta_{2,d}=\gamma_{2,d}+d(d+2)$. 
It turns out that it is the sum
of three terms: the univariate kurtosis index of $Y_1^*$, whose
expression is given in Proposition~\ref{p:sm-sn-g1g2}, the
kurtosis index $\beta_{2,d-1}$ of the $(d-1)$-dimensional scale
mixture of normal distribution $(Y_2^*, \ldots, Y_d^*)\T$, which
is given by $(d-1)(d+1)E(S^{2})^{-2}E(S^{4})$, and a term which is related 
with the fourth moment of $Y*$ through  
$\mu_{1,1,i,i}$, for any $i \in \{2,\ldots,d\}$.

When $Y \sim ST_d(\xi, \Omega,\alpha,\nu)$ explicit expressions of
the two indices can be easily obtained taking into account the
well known result
\[
E(S^{m/2})=\frac{(\nu/2)^{m/2}\Gamma((\nu-m)/2)}{\Gamma(\nu/2)},
\]
leading to
\begin{eqnarray*}\label{f:g-1-mst}
\gamma_{1,d}&=&(\gamma_1^*)^2 +
3(d-1)\frac{\mu_*^2}{(\nu-3)\sigma_*^2},  \qquad \mathrm{if} \, \nu >3,\\
\gamma_{2,d}&=&\beta_2^* +
(d^2-1)\frac{(\nu-2)}{(\nu-4)}+\frac{2(d-1)}{\sigma_*^2}\left[\frac{\nu}{\nu-4}-\frac{(\nu-1)\mu_*^2}{\nu-3}
\right] -d(d+2),\\
 & & \mathrm{if} \, \nu>4
\end{eqnarray*}
where
\[
\mu_*=\delta_*\left(\frac{\nu}{\pi}
\right)^{1/2}\frac{\Gamma((\nu-1)/2)}{\Gamma(\nu/2)}, \qquad
\sigma_*^2= \frac{\nu}{\nu-2}-\mu_*^2,
\]
and the explicit expressions of $\gamma_1^*$ and
$\gamma_2^*=\beta_2^* -3$ are given in Azzalini and Capitanio
(2003, p.\,382). 

Note that an equivalent expression, obtained through a different method, for $\beta_{2,d}=\gamma_{2,d}+d(d+2)$, is given in Kim and Mallik (2009). Finally, note also that the expression of $\gamma_{1,d}$ and $\gamma_{2,d}$ given in Proposition~\ref{p:mardia-d} reduces to the corresponding ones for the skew-normal distribution when $S$ is such that $\mathrm{pr}(S=1)=1$, while $\gamma_{2,d}$ is the index of multivariate kurtosis of a scale mixture of normal distributions with mixing variable $S$ when $\delta_* = 0$.

The following proposition provides a further example of a pair of scatter matrices that can be used for obtaining the linear transform to convert a scale mixture of skew-normal variates into a canonical form. The proof of the proposition contains the proof of the fact that if two scatter matrices are diagonal when the considered variate is in canonical form, then it is expected that by applying to them the procedure described in  Proposition~\ref{p:sn-ics} we obtain a matrix $H$ that induces a canonical form.
\begin{proposition} \label{p:sn-ics-sigmakappa}
Consider the scale mixture of skew-normal distribution $Y = \xi +
\omega S Z$, where $Z \sim \SN_d(0, \bar \Omega, \alpha)$, and define the scatter matrix 
\[
  \mathcal{K} = E\left\{\left[(Y-\mu)\T\Sigma\inv(Y-\mu)\right]^2 (Y-\mu)(Y-\mu)\T\right\}. 
\]
Let $M' = \Sigma^{-1/2}\mathcal{K}\Sigma^{-1/2}$, where $\Sigma^{1/2}$ is the unique positive definite symmetric square root of $\Sigma$,  and $\Sigma$ is the covariance matrix of $Y$. Let $Q' \Lambda' {Q'}\T$ denote the  spectral decomposition of $M'$. Then the transform 
\[
Y^* = H\T(Y-\xi),
\]
where $H=\Sigma^{-1/2}Q'$, converts $Y$ into a canonical form. 
\end{proposition}
\emph{Proof.} By means of the results contained in Proposition~\ref{p:cf-properties} it is possible to show that when a scale mixture of skew-normal distribution is in canonical form, then both the scatter matrices $\mathcal{K}$ and $\Sigma$ are diagonal. Let $\mathcal{K^*}=\tilde{H}\T\mathcal{K}\tilde{H} $ and $\Sigma^*=\tilde{H}\T \Sigma \tilde{H}\T$ denote such matrices, where $\tilde{H}$ is a matrix such that $\tilde{H}\T(Y-\xi)$ is in canonical form. The equality $M'q'_j =\lambda'_j q'_j$, where $q'_j$ is the $j$-th column of the matrix $Q'$ and $\lambda'_j=\Lambda'_{jj}$ is the corresponding eigenvalue, implies that the equality $\Sigma^{*-1}\mathcal{K^*}\tilde{H}^{-1}(\Sigma^{-1/2}q'_j)=\lambda'_j \tilde{H}^{-1}(\Sigma^{-1/2}q'_j)$ must also hold true; since both $\mathcal{K^*}$ and $\Sigma^*$ are diagonal, the equality is fulfilled when all the eigenvalues of $M'$ are equal, or when $\tilde{H}^{-1}(\Sigma^{-1/2}Q')\propto I_d$. The first circumstance is out of interest, because it would imply that we are considering two scatter matrices which are proportional, the second one implies that the columns of $\Sigma^{-1/2}Q'$ are proportional to the corresponding columns of $\tilde{H}$, and the proposition is proved. QED

On the basis of Propositions~\ref{p:sn-ics} and~\ref{p:sn-ics-sigmakappa} we see that the matrix $H$ that defines the canonical form can be obtained working with the pair $(\Omega, \Sigma)$ or with $(\Sigma, \mathcal{K})$, no matter which one between them. However it is important to highlight the auxiliary information given by  this technique, which essentially relies on a spectral decomposition. In particular, it is straightforward to note that the trace of the matrix $\Omega^{-1}\Sigma$, or equivalently, of $M$, is equal to the sum of the variances of the marginal univariate components of the canonical form, while the trace of the matrix $\Sigma^{-1}\mathcal{K}$, or equivalently, of $M'$, is equal to $\beta_{2,d}$.
\section{The mode of the multivariate skew-normal and skew t
distributions} 
The mode of the skew-normal and skew $t$
distributions cannot be calculated in closed form, so one needs to
resort to numerical methods. In this section it is proved the uniqueness of the mode in the $d$-dimensional case, and it is
shown that its computation can be
reduced to an equivalent one-dimensional problem, drastically
reducing the dimensionality of the original problem. From the expression 
of the mode which is obtained, it also turns out that  
the mode, the mean and the location parameter are aligned. More specifically, they lie in a one dimensional linear manifold 
of direction $\omega \delta$. Thus, the departure from symmetry of these distributions is characterized by a displacement of the probability mass along this direction. 
The above issues are briefly  discussed also for the general case of scale mixture of skew-normal
distributions. 

For later use, we recall that the density function
of a $d$-dimensional skew $t$ variate as given by Azzalini and
Capitanio (2003, expression 26) is
\begin{equation}   \label{f:mst-dens}
f_Y(y) = 2 \, t_d(y - \xi ; \nu) T_1 \left\{\alpha\T \omega \inv
(y - \xi)\left(\frac{\nu + d}{Q_y + \nu} \right)^{1/2}; \nu + d 
\right\} \qquad (y\in \Real^d),
\end{equation}
where $Q_y = (y-\xi)\T \Omega\inv (y-\xi)$, $t_d(x ; \nu)$ is the
density function of a $d$-dimensional $t$-variate with $\nu$
degrees of freedom, $T_1(x;\nu+d)$ is the scalar $t$ distribution
function with $\nu+d$ degrees of freedom. A random variable having
density (\ref{f:mst-dens}) will be denoted by
$ST_d(\xi,\Omega,\alpha,\nu)$.
\begin{proposition}  \label{f:dsn-mode}
Let $Z \sim SN_d(\xi, \Omega, \alpha)$. Then the unique mode of $Z$ is
\[
 M_0 = \xi + \frac{m_0^*}{\alpha_*}\omega \bar\Omega \alpha =  \xi + \frac{m_0^*}{\delta_*}\omega
 \delta
\]
where $m_0^*$ is the mode of a scalar $SN_1(0,1,\alpha_*)$ random
variable.

\end{proposition}
\emph{Proof.} Consider first the mode of the canonical form $Z^*
\sim SN_d(0, I_d, \alpha_{Z^*})$. If we calculate the mode by
imposing the gradient of the density function to be equal to the
null vector, the system of equations to
be solved turns out to be
\begin{eqnarray*}\label{f:grad-sn}
 z_1\Phi(\alpha_* z_1)-\phi_1(\alpha_* z_1)\alpha_*&=& 0 \\
 z_2 \Phi(\alpha_* z_1)&=& 0 \\
 \vdots &=& \vdots\\
z_d \Phi(\alpha_* z_1)&=& 0,
\end{eqnarray*}
where $z_i$, $i=1,2,\ldots,d$ denotes the $i$th entry of the vector $z\in \Real^d$. The last $d-1$ equations are satisfied when $z_i=0$ for
$i=2,\ldots,d$, whilst the unique root (for the uniqueness see
Azzalini, 1985, Property D) of the first one corresponds to the
mode, say $m_0^*$, of a $SN_1(0,1,\alpha_*)$, so that the mode of
$Z^*$ is the vector $M_0^*=(m_0^*, 0, \ldots,0)\T =
(m_0^*/\alpha_*)\alpha_{Z^*}\T$. Recalling that $Z=\xi +
\omega C\T P Z^*$ and $\alpha_Z^* = P\T C \alpha$, and taking into
account that the mode is equivariant with respect to affine
transformations, the mode of $Z$ turns out to be
\[
 M_0 = \xi + \frac{m_0^*}{\alpha_*} \omega C\T P P\T C \alpha=
     \xi + \frac{m_0^*}{\alpha_*}\omega \bar\Omega \alpha = \xi + \frac{m_0^*}{\delta_*}\omega
     \delta,
\]
where the last equality follows taking into account
(\ref{f:delta}) and (\ref{f:alpha}). QED

\begin{proposition}  \label{f:dst-mode}
Let $Y \sim ST_d(\xi, \Omega, \alpha, \nu)$. Then the unique mode of $Y$
is
\[
 M_0 = \xi + \frac{y_0^*}{\alpha_*}\omega \bar\Omega \alpha = \xi + \frac{y_0^*}{\delta_*}\omega
 \delta
\]
where $y_0^*\in \Real$ is the unique solution of the equation
\[
y(\nu+d)^{1/2}T_1(w(y); \nu+d) - t_1(w(y); \nu+d)\nu \alpha_*
(\nu+y^2)^{-1/2}=0,
\]
where $w(y)=\alpha_* y \left(\dfrac{\nu+d}{\nu+y^2}\right)^{1/2}$.
\end{proposition}
\emph{Proof.} As for the skew-normal case, the canonical form $Y^*
\sim ST_d(0, I_d, \alpha_{Y^*}, \nu)$, where
$\alpha_{Y^*}=(\alpha_*, 0, \ldots,0)\T$ is considered, and the
mode is calculated by imposing the gradient of the density
function to be equal to the null vector. The system of equations
to solve turns out to be
\begin{eqnarray*}\label{f:grad-st}
 x_1 T_1(\alpha_* x_1 c(x); \nu+d)-
         \frac{t_1(\alpha_* x_1c(x); \nu+d)}{(\nu+x\T x)^{1/2}(\nu+d)^{1/2}}(\nu+x\T x-x_1^2)\alpha_* & \hspace{-3pt} =& 0\\
 x_2 \left[T_1(\alpha_* x_1 c(x); \nu+d)+
         \frac{t_1(\alpha_* x_1c(x); \nu+d)}{(\nu+x\T x)^{1/2}(\nu+d)^{1/2}}x_1\alpha_*\right]&=& 0\\
\vdots &=& \vdots\\
 x_d \left[T_1(\alpha_* x_1 c(x); \nu+d)+
         \frac{t_1(\alpha_* x_1 c(x); \nu+d)}{(\nu+x\T x)^{1/2}(\nu+d)^{1/2}}x_1\alpha_*\right]&=& 0,
\end{eqnarray*}
where $x=(x_1,x_2,\ldots,x_d)\T$ and
$c(x)=\{(\nu+d)/(\nu+x\T x)\}^{1/2}$. 
First note that
the function on the left hand side of the first equation can be
equal to zero only if $x_1\ge 0$. This fact implies that the
remaining equations are equal to zero if and only if $x_i=0$ for
$i = 2,\ldots, d$. Hence the mode of $Y^*$ is
$M_0^*=(y_0^*,0,\ldots,0)\T$, where the scalar value $y_0^*\ge 0$
is the solution of
\begin{equation} \label{f:uniq-st}
y T_1(w(y);\nu+d)-
\frac{t_1(w(y);\nu+d)}{(\nu+y^2)^{1/2}(\nu+d)^{1/2}}\nu\alpha_*=0, 
\end{equation}
where $w(y)=\alpha_* y \left(\dfrac{\nu+d}{\nu+y^2}\right)^{1/2}$. To see that equation (\ref{f:uniq-st}) admits a unique
solution, first notice that when $y_0^*\ge 0$ the function on the
right hand side is the difference between a strictly increasing
function and a strictly decreasing one. Furthermore, when
$y_0^*=0$ the latter is greater than zero while the former is
equal to zero, and as $y_0^* \rightarrow \infty$ the latter goes
to zero while the former goes to $\infty$. Hence, there exists a
unique point in which their difference is equal to zero. The
expression of the mode of $Y$ is obtained on the basis of
arguments analogous to those used for the mode of a
multivariate skew-normal distribution. QED

Note that a different proof for the uniqueness of the mode for the multivariate skew \emph{t} distribution has been independently developed by Azzalini and Regoli (2012).

The issue of finding the mode of other members of the family of
scale mixture of skew-normal distributions can be tackled in a
similar way. An open problem, which is not investigated here, is to
assess the uniqueness of the solution. 

It is straightforward to see that if a point of $\Real^d$ is the mode of the canonical form of a
$d$-dimensional skew scale mixture of skew-normal variates, then it should be of type $(y_0^*,0,\ldots,0)\T$, where  the real number $y_0^*$
is such that
\[
\int_0^{\infty}s^{-d-1} \phi\left(\frac{y_0^*}{s}\right)
\left\{\frac{y_0^*}{s}\Phi\left(\alpha_*
\frac{y_0^*}{s}\right)-\alpha_* \phi\left(\alpha_*
\frac{y_0^*}{s}\right)\right\}f_S(s)\d s = 0,
\]
where $f_S(s)$ denotes the density function of $S$.
This implies that, as for the skew-normal and skew $t$
distributions, the mode of a scale mixture of skew-normal distributions will be
of the form 
\[
\xi + \dfrac{y_0^*}{\delta_*} \omega \delta.
\]
\subsection*{Acknowledgements}
The author is grateful to Adelchi Azzalini for helpful and
stimulating discussions. This research has been supported by the
grant scheme PRIN 2006, grant No. 2006132978, from MIUR, Italy.

\subsection*{References}

\biblioitem
  Adcock, C.~J. (2009). Asset pricing and portfolio selection based 
on the multivariate extended skew-Student-$t$ distribution.   
\emph{Ann. Oper. Res.} In press.
  
\biblioitem
  Azzalini, A. (1985). A class of distribution which includes the normal
  ones.   \emph{Scand. J. Statist.} {\bf 12}, 171--178.

\biblioitem
  Azzalini, A. \& Capitanio, A. (1999).  Statistical
  applications of the multivariate skew normal distribution.  \emph{J.
  Roy. Statist. Soc., B} \textbf{61} 579--602.

\biblioitem
  Azzalini, A. \& Capitanio, A. (2003).  Distributions generated by perturbation of symmetry
  with emphasis on a multivariate skew $t$-distribution.  \emph{J.
  Roy. Statist. Soc., B} \textbf{65} 367--389.

\biblioitem
  Azzalini, A. \& Dalla Valle, A. (1996).  The multivariate
  skew normal distribution.  \emph{Biometrika} \textbf{83}, 715--726.
  
\biblioitem
  Azzalini, A. \& Genton, M.~G. (2008).  Robust likelihood methods based on the skew-t and related distributions.  \emph{Int. Statist. Rev.} \textbf{76}, 106--129.

\biblioitem
Azzalini, A. \& Regoli, G. (2012).  Some properties of skew-symmetric distributions.  \emph{Annals of the Institute of Statistical Mathematics} \textbf{64}, 857--879.

\biblioitem
  Branco, M.~D.\ \& Dey, D.~K.\ (2001).
  A general class of multivariate skew elliptical distributions.
  \emph{Journal of Multivariate Analysis} \textbf{79}, 99--113.

\biblioitem
Wang, J., Genton, M.~G., (2006).
  The multivariate skew-slash distribution.
  \emph{J. Statist. Plann. Inference} \textbf{136}, 209--220.
  
\biblioitem
Genton, M.~G.,  Li, H. \& Xiangwei, L. (2001).
  Moments of skew normal random vectors  and  their quadratic forms.
  \emph{Statist. \& Prob.\ Lett.} \textbf{51}, 319--325.
  
\biblioitem
Kim, H.~M. (2008).
 A note on scale mixtures of skew normal distribution.
  \emph{Statist. \& Prob.\ Lett.} \textbf{78}, 1694--1701.

\biblioitem
Kim, H.~M., \& Mallik, B.~K. (2009).
  Corrigendum to: "Moments of random vectors with skew $t$ distribution and their quadratic forms"[Statist. Probab. Lett. 63 (2003) 417--423].
  \emph{Statist. \& Prob.\ Lett.} \textbf{79}, 2098--2099.

\biblioitem
Mardia, K.V. (1970).
  Measures of multivariate skewness and kurtosis with
  applications.
  \emph{Biometrika} \textbf{57}, 519--530.
  
\biblioitem
Mardia, K.V. (1974).
  Applications of some measures of multivariate skewness and kurtosis
  in testing normality and robustness studies.
  \emph{Sankhya B} \textbf{36}, 115--128.

\biblioitem
Meucci, A. (2006).
  Beyond Black-Litterman: views on non-normal markets.
  \emph{Risk Magazine} \textbf{19(2)}, 87--92.
 
\biblioitem
Tyler, D.~E., Critchley, F., D\"umbgen, L., \& Oja, H. (2009).
  Invariant co-ordinate selection (with discussion).
   \emph{J. Roy. Statist. Soc., B} \textbf{71}, 549--692.

\biblioitem
Walls, W.~D. (2005).
  Modeling heavy tails and skewness in film returns.
   \emph{Applied Financial Economics} \textbf{15(17)}, 1181--1188.

\end{document}